\documentclass[10pt]{article}
\usepackage{mathrsfs}
\usepackage{multicol}
\usepackage{amsmath}
\usepackage{graphicx}
\usepackage{float}
\makeatletter
\newcommand\figcaption{\def\@captype{figure}\caption}
\makeatother
\usepackage{flafter}
\usepackage{amsmath}
\usepackage{latexsym}
\usepackage{feynmf}
\usepackage{cite}
\usepackage{bm}
\usepackage{amsfonts}
\usepackage{indentfirst}
\usepackage{balance}
\newcommand{\ct}[1]{{\textsuperscript{{\cite{#1}}}}}

\newcommand{\bee}{\begin{equation}}
\newcommand{\ee}{\end{equation}}
\newcommand{\beea}{\begin{eqnarray}}
\newcommand{\eea}{\end{eqnarray}}

\newcommand{\kk}{{\bf k}}

\newcommand{\vv}[1]{{\mathbf #1}}
\newcommand{\ce}{\textrm{e}}
\newcommand{\ci}{\textrm{i}}
 \newlength{\halfpagewidth}
 \divide\halfpagewidth by 2

\newcommand{\sd}[1]{#1 \!\!\!/}
\newcommand{\sdd}[1]{#1 \!\!\!/}
\begin{document}
\title{ Anomalous Valley Magnetic Moment of Graphene}
\author{{Daqing Liu, Shengli Zhang, Erhu Zhang, Ning Ma, Huawei Chen}
\\
        {\small Department of Applied Physics, Non-equilibrium Condensed Matter and Quantum
        Engineering Laboratory,  }\\
        {\small Key Laboratory of Ministry of
        Education, Xi'an Jiaotong University, Xi'an, 710049, China}\\
       }
       \date{}
\maketitle

 \abstract{ Carrier interactions on graphene are studied.
The study shows that besides the well known Coulomb repulsion
between carriers, there also exist four-fermion interactions
associated with U-process, one of which attracts carriers in
different valleys. We then calculate the contributions to valley
magnetic moment from vertex correction and from four-fermion
corrections explicitly. The relative contributions are -18\% and 3\%
respectively. At last we point out that we can mimic heavy
quarkonium system by carrier interactions in graphene.}

\begin{multicols}{2}

\section{Introduction}

Graphene \cite{discovery}, newly discovered two-dimensional
crystals, has attracted more and more attentions of theorists and
experimentalists\ct{back,back2,back3,ijmp}. In graphene, there is a
typical valley degeneracy, corresponding to the presence of two
different valleys in the band structure. However, as stated in the
reference \cite{ab-effect}, such degeneracy makes it difficult to
observe the intrinsic physics of a single valley in
experiments\ct{fiction,berry}. How to distinguish carriers in the
two valleys is therefore always a topic attracting
literatures\ct{ab-effect,prb77,nat2007,niu}.

Ref. \cite{niu} pointed out that in close analogy with the spin
degree, there is an intrinsic magnetic moment associated with the
valley index, which was called as valley magnetic moment (VMM).
At tree level the valley magnetic moment is about 30 times that of
the usual spin magnetic moment, therefore, "valleytronics" provides
a new and much more standard pathway to potential applications in a
broad class of semiconductors as compared with the novel valley
device in graphene nanoribbon\ct{nat2007}. However, since in
graphene the effective coupling $\frac{e^2}{\varepsilon\hbar v}\sim
1$, a question to be posed is to what extent the calculation in ref.
\cite{niu} is valid.

To answer the question, we first study carrier interactions. The
study shows that, in tight binding approximation, besides the
well-known Coulomb repulsion between electrons, there are also
four-fermion interactions associated with U-process. The
four-fermion interactions are type dependent and more significant,
one of them attracts carriers in different valleys. Armed with the
understanding of the interactions, we point out that there are two
corrections to VMM at one-loop level. One is the vertex correction
and the other is the four-fermion correction. The vertex correction
is similar to the anomalous magnetic moment of a particle in quantum
electrodynamics (QED) except that carrier interactions on graphene
are not "Lorentz covariant". Therefore, such correction always
appears even in a one-valley system. Meanwhile, the valley degree is
similar to the flavor degree in particle physics or high-energy
physics. To compute anomalous magnetic moment of a particle due to
flavor degree, one should also consider the weak interaction, an
interaction between flavor degree. Our Yakawa-like four-fermion
interactions are similar to the lower-energy approximation of the
weak interactions. In this way, the correction to VMM due to valley
degree appears at one-loop level. In contrast, such correction can
not occur in QED.

Our study shows that the total correction is about $-15\%$.
Furthermore, since the corrections are independent on the divergence
of the loop calculations, VMM can be used to check the validity of
the perturbational calculation.

\section {Carrier interactions}
Here we study carrier interactions. The study shows that besides the
well known Coulomb repulsion, there are also four-fermion
interactions between carriers at different valleys, which are not
only short-range but also contacting ones.

For simplicity, we set $\hbar\equiv 1$ and
$X(\vv{r}-\vv{r}_{A^\prime})$ the normalized orbital $p_z$ wave
function of electron bound to atom $A^\prime$, {\it i.e.} it
satisfies $\int
d\vv{r}X(\vv{r}-\vv{r}_{A^\prime})X(\vv{r}-\vv{r}_{A^{\prime\prime}})
=\delta_{A^\prime A^{\prime\prime}}$ \ct{wallace}. A-electron wave
function $\psi_A(\vv{k})$ in position space is
$\psi_A(\vv{k})=\sqrt{\omega}/ (2\pi)\sum\limits_A \ce^{\ci
\vv{k\cdot\vv{r}_A}}X(\vv{r}-\vv{r}_A)$, where $\omega=\sqrt{3}
a^2/2$ is the area of the hexagonal cell. For B-electron the case is
similar. We then have
$<\psi_{A^\prime_0}(\vv{k}^\prime)|\psi_{A_0}(\vv{k})>=\delta_{A_0
A_0^\prime}\delta(\vv{k}-\vv{k^\prime})$, where $A_0, A_0^\prime=A$
or $B$.

To study carrier interactions, we consider \bee
V(\kk)=\frac{(2\pi)^4}{N\omega}
\psi_{A_2^\prime}^*(\kk_2+\kk)\psi_{A_2}(\kk_2)\hat{V}
\psi_{A_1^\prime}^*(\kk_1-\kk)\psi_{A_1}(\kk_1), \ee where $A_1,\,
A_1^\prime,\,A_2,\, A_2^\prime$ equal to $A$ or $B$. If we ignore
interchange interactions, the main contribution to
$X^*(\vv{r}^\prime-\vv{r_{A_2^\prime}})X(\vv{r}^\prime-\vv{r_{A_2}})
\hat{V}X^*(\vv{r}-\vv{r_{A_1^\prime}})X(\vv{r}-\vv{r_{A_1}})$ should
be at vicinity $\vv{r}_{A_1^\prime}=\vv{r}_{A_1}$,
$\vv{r}_{A_2^\prime}=\vv{r}_{A_2}$, $\vv{r}^\prime\approx
\vv{r}_{A_2}$ and $\vv{r}\approx \vv{r}_{A_1}$. We get \beea
V(\kk)&=&\delta_{A_1A_1^\prime}\delta_{A_2A_2^\prime}\sum\limits_{A_1A_2}
e^{\ci \kk(\vv{r}_{A_1}-\vv{r}_{A_2})}\frac{\omega e^2}{
N|\vv{r}_{A_1}-\vv{r}_{A_2}|} \nonumber \\
&=&\delta_{A_1A_1^\prime}\delta_{A_2A_2^\prime}\sum\limits_{A_2}
e^{\ci \kk(\vv{r}_{A_1}-\vv{r}_{A_2})}\frac{\omega e^2}{
|\vv{r}_{A_1}-\vv{r}_{A_2}|}, \label{ints} \eea where we have fixed
$A_1$ in the last step. We shall ignore the two delta functions
thereinafter.

\begin{center}
%~\\[\intextsep]
\begin{minipage}{0.45\textwidth}
\centering
%\vskip 0.2in
\includegraphics[width=2in]{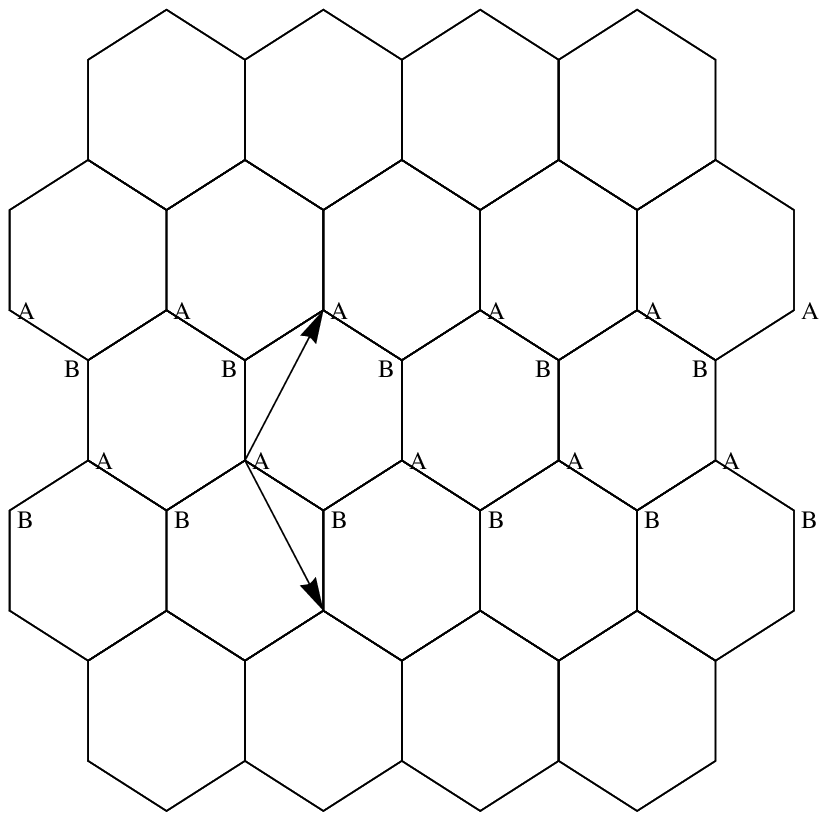}%
\vskip 1.9in \figcaption{Graphene hexagonal lattice constructed as a
superposition of two triangular lattices A and B, with bases vectors
$\vv{a}_1=a(\frac{1}{ 2},\frac{\sqrt{3}}{2})$ and
$\vv{a}_2=a(\frac{1}{ 2},-\frac{\sqrt{3}}{ 2})$, where lattice
constant $a=|\vv{a}_1|=|\vv{a}_2|$. The reciprocal lattice vectors
are $\vv{b}_1=\frac{2\pi}{ a}(1,\frac{1}{ \sqrt{3}})$ and
$\vv{b}_2=\frac{2\pi}{ a}(1,-\frac{1}{ \sqrt{3}})$ respectively. }
\label{lattice}
\end{minipage}
\setlength{\intextsep}{0in plus 0in minus 0.2in}
%\\[\intextsep]
\end{center}

Without loss of generality, we set $A_1=A$. To compute interactions
between carriers, we first mark coordinates of $A$ and $B$ with two
integers $n_1$ and $n_2$. From Fig.\ref{lattice}, the coordinate of
one atom A is set as $(0,0)$. Then, for infinitely large graphene,
coordinates of atom $A$ are depicted as $(n_1/2,\,
\sqrt{3}(n_1-2n_2)/2)a$ and coordinates of atoms $B$ $(n_1/2,\,
\sqrt{3}(n_1-2n_2+\frac{2}{3})/2)a$ respectively, where $a$ is
lattice constant and $n_1$, $n_2$ are arbitrary integers.

We put our focus on the interactions between electrons around $\pm
{\bf K}=\pm (4\pi/ 3a,0)$. We first study the case where there is no
valley-valley transition during interactions. For this case, we
suppose $|\vv{k} a|\ll 1$. In Eq. (\ref{ints}) the function in the
summation is a slow-moving function, therefore, the summation can be
replaced by an integral, \addtocounter{equation}{1}
\begin{align}
V_c(\kk)\approx e^2\int d\vv{r} \frac{e^{\ci \kk\cdot\vv{r}}}{
|\vv{r}|}= \frac{2\pi e^2 }{ \varepsilon k}. \tag{\theequation
a}\label{vc}
\end{align}
where we have inserted the effective permittivity
$\varepsilon$ in the last equation to include screening effect. We
thus obtain the well known Coulomb interaction. The type of $A_2$
does not influence the results, that is, the Coulomb repulsion works
both for carriers in the same valley and for carriers in the
different valleys.

Besides the well known Coulomb interactions, there are other
interactions which is related to valley-valley transition. Such
interactions correspond to a U-process and therefore $\vv{k}\approx
\pm ({4\pi/ 3a},0)$. To deal with such case, we substitute
$\vv{k}+(\frac{4\pi}{ 3a},0)$ for $\vv{k}$ in Eq. (\ref{ints}).

We first consider A-B interactions, that is, $A_2=B$ in Eq.
(\ref{ints}). We get then
\begin{align}
 V_d(\vv{k})\approx e^2\sum\limits_{n_1\neq 0}\frac{a}{2}e^{-\ci
k_x{n_1 a\over 2}-\ci{2\pi\over 3}n_1} 2K_0(|k_y \frac{n_1 a}{2}|) \nonumber \\
 +e^2\frac{a}{2}
\sum\limits_{n_2}\frac{e^{\ci k_y \sqrt{3}a(n_2-\frac{1}{3})
}}{|n_2-\frac{1}{3}|}
%\nonumber \\ &\approx &
\approx 0.1 e^2a.\tag{\theequation b} \label{v1} \end{align}
Compared to the long-wavelength result in Eq. (\ref{vc}), the
valley-valley interaction suffers a coefficient suppression due to
the large momentum transfer. However, since such valley-valley
interactions are short-range, it is not needed to consider screening
effect. We therefore does not insert the effective
permittivity $\varepsilon$ in the above equation.

Whereas when $A_2=A$, one should subtract the contribution from self
interactions, which corresponds to $(n_1,n_2)=(0,0)$ in Eq.
(\ref{ints}). The result is then \begin{align} V_s(\vv{k})\approx
-1.55 e^2a. \tag{\theequation c} \label{v2}\end{align} Here, the
large negative coefficient $-1.55$ is due to the subtraction.

Since Coulomb interaction is long-range, it does not depend on the
distributing detail of the adjoint electrons. Thus, as shown in Eq.
(\ref{vc}), such interaction is type-independent. However, the
four-fermion form of the U-process implies that such interactions
are short-range and they therefore depend on the distributing
detail. Therefore, as shown by  (\ref{v1}) and (\ref{v2}), such
interactions are type-dependent. Reference
\cite{prb77-115410,czhang} also proposed four-fermion interactions
from different aspects. In Ref. \cite{prb77-115410} the
authors add a near-neighbor interaction term and then, when they
carry out momentum integral in the first Brillouin zone, they adhere
the short-range interaction with the usual Coulomb interaction at
$|\vv{k}|=\frac{1}{2}\frac{\pi}{\sqrt{3}a}$, where $\vv{k}$ is the
transfer momentum. In contrast, in our approach, there is no
artificial adhering and the interactions due to the valley transition
are shown explicitly. Furthermore, the results obtained by our
approach are suitable to take the quantum field theory
calculations.

%\revision{ The authors investigate different types of carrier
%interactions in graphene. On the
% one hand, for small momentum transfer $|ka|<<1$, they obtain the usual Coulomb
% interaction. On the other hand, for large momentum transfer (intervalley coupling),
% the claim is that they find a new type of interaction (a U-process). It is not clear
% to me to what extend this manuscript goes beyond existing work on a similar topic,
% e.g. Ref.[19]. The authors should clearly mention in which aspect their work is new
% compared to previous analysis of carrier interaction in graphene.}

We emphasize that, besides the vertex correction, $V_s$ also
contributes to VMM. Furthermore, since $V_s<0$, it takes attracting
force between electrons in different valleys. The interaction may
play crucial role in superconduction phenomena \ct{supercurrent}.
Therefore, $V_s$ deserves further research.

\section {The formal development of Lagrangian}
We first define two two-component spinors $\varphi$ and $\chi$ as
follows: $\varphi=\left(
                                              \begin{array}{c}
                                                a_{\bf K}(\bf p) \\
                                                b_{\bf K}(\bf p) \\
                                              \end{array}
                                            \right)$ and $\chi=\left(
                                              \begin{array}{c}
                                                -\ci b_  {\bf -K}(\bf p) \\
                                               \ci a_{\bf -K}(\bf p) \\
                                              \end{array}
                                            \right)$, where ${\bf
                                            \pm
                                            K}$
are two valleys. To describe the graphene dynamics in field theory
language, we read the Lagrangian,
 \beea
\label{bare-lagrangian} \mathcal{L}_0 &=&
\bar{\varphi}(\ci\gamma^0{\partial_t}+\ci v{\bf \gamma}\cdot
\nabla-m)\varphi-e\bar{\varphi}(\gamma^0A^0-\beta\vv{\gamma}\cdot
\vv{A})\varphi  \nonumber
\\ && + \bar{\chi}(\ci\gamma^0{\partial_t} +\ci
v\mbox{\textbf{$\gamma$}}\cdot \nabla+m)\chi
   -e\bar{\chi}(\gamma^0A^0-\beta\vv{\gamma}\cdot \vv{A})\chi \nonumber\\
   && -\frac{\lambda_1}{2} \bar{\varphi}\gamma^1\chi\bar{\chi}\gamma_1\varphi
   -\frac{\lambda_2}{2} \bar{\varphi}\gamma^2\chi\bar{\chi}\gamma_2\varphi,\eea
where $\beta={v/ c}$, $v$ is the Fermi velocity of carriers, $c$ is
the effective light velocity in graphene, $\lambda_1=-(V_s-V_d)/2 $
and $\lambda_2=-(V_s+V_d)/2$. We also set three gamma matrices as
$\gamma^0=\gamma_0=\sigma_3=\left(
                         \begin{array}{cc}
                           1 & 0 \\
                           0 & -1 \\
                         \end{array}
                       \right)$, $\gamma^1=g^{11}\gamma_1=-\gamma_1=\gamma^0\sigma_1=\left(
                         \begin{array}{cc}
                           0 & 1 \\
                           -1 & 0 \\
                         \end{array}
                       \right)$ and $\gamma^2=g^{22}\gamma_2=-\gamma_2=\gamma^0\sigma_2=\left(
                         \begin{array}{cc}
                           0 & -i \\
                           -i & 0 \\
                         \end{array}
                       \right)$, where $\sigma_i$'s are three Pauli matrices and
metric matrix $g^{\mu\nu}=diag\{1,-1,-1\}$. Since four-fermion
interactions in Eqs. (\ref{v1}) and (\ref{v2}) are contacting ones,
it is not necessary to introduce corresponding gauge field.
In the above equation the energy gap $m$ can be used to
improve the use of graphene in making transistors and is therefore
the one of the hot spots of literatures. In Ref. \cite{mass}, the
authors investigate the energy gap of graphene on a substrate BN,
which is generated by the breaking of the A-B sublattice symmetry.
However, such energy gap has not been observed up to now. In Ref.
\cite{gap} the authors report that single layer graphene on SiC has
a gap of 0.26eV, but the result is under debate \cite{ohta,zhou2}.

Utilizing the definitions of $\gamma$ matrix and boost generators
corresponding to "Lorentz" transformation we find positive solution
and negative solution for $\varphi$ field (or negative solution and
positive solution for $\chi$ field) as
$u(\vv{p})=(\sqrt{p^0+m},{v(p^1+\ci p^2)/ \sqrt{p^0+m}})^T$ and
$v(\vv{p})=({v(p^1-\ci p^2)/ \sqrt{p^0+m}},\sqrt{p^0+m})^T$
respectively. They meet $ \sd{\tilde{p}}u(p)=mu(p), \,
\bar{v}(p^\prime)\sd{\tilde{p}}^\prime=-\bar{v}(p^\prime)m$, where
$\sdd{\tilde{p}}=\gamma_\mu\tilde{p}^\mu$ and
$\tilde{p}=(p^0,\,v\vv{p})$.

However, the Lagrangian in Eq. (\ref{bare-lagrangian}) is a bare one
and it needs renormalization to match the observable
quantities\ct{ffi}. Having set
$\varphi=Z^{1/2}_2\varphi_r,\,\chi=Z^{1/2}_2\chi_r$ and
$A=Z^{1/2}_3A_r$, where $\varphi_r$, $\chi_r$ and $A_r$ are
renormalized quantities, we split each term of the Lagrangian
into two pieces as follows:
 \beea
\label{lagrangian} \mathcal{L} &=&
\bar{\varphi}_r(\ci\gamma^0{\partial_t} +\ci v_r{\bf \gamma}\cdot
\nabla-m_r)\varphi_r \nonumber \\ &&
-e_r\bar{\varphi}_r(\gamma^0A^0_r-\beta_r\vv{\gamma}\cdot
\vv{A}_r)\varphi_r  \nonumber \\ &&+
\bar{\chi}_r(\ci\gamma^0{\partial_t}+\ci v_r\gamma\cdot
\nabla+m_r)\chi_r
   \nonumber\\
   &&  -e_r\bar{\chi}_r(\gamma^0A^0_r-\beta_r\vv{\gamma}\cdot
   \vv{A}_r)\chi_r\nonumber \\ &&
   -\frac{\lambda_{1r}}{2}
   \bar{\varphi}_r\gamma^1\chi_r\bar{\chi}_r\gamma_1\varphi_r
   -\frac{\lambda_{2r}}{2}
   \bar{\varphi}_r\gamma^2\chi_r\bar{\chi}_r\gamma_2\varphi_r
   \nonumber \\ &&
   +\bar{\varphi}_r(\ci\delta_2\gamma^0\partial_t+i\delta_v
   v_r\vv{\gamma}\cdot\nabla-\delta_m)\varphi_r \nonumber \\ &&
   +\bar{\chi}_r(\ci\delta_2\gamma^0\partial_t+i\delta_v
   v_r\vv{\gamma}\cdot\nabla+\delta_m)\chi_r \nonumber \\ &&
   -e_r\bar{\varphi}_r(\delta_1\gamma^0A^0_r-
   \beta_r\delta_c\gamma\cdot\vv{A}_r)\varphi_r \nonumber \\ &&
   -e_r\bar{\chi}_r(\delta_1\gamma^0A^0_r-\beta_r\delta_c\gamma\cdot\vv{A}_r)\chi_r
   \nonumber \\ &&-\frac{\delta_{1\lambda}}{2} \bar{\varphi}_r\gamma^1\chi_r\bar{\chi}_r\gamma_1\varphi_r
   -\frac{\delta_{2\lambda}}{2}
   \bar{\varphi}_r\gamma^2\chi_r\bar{\chi}_r\gamma_2\varphi_r, \eea
where counterterm coefficients $\delta_2=Z_2-1$,
$\delta_m=Z_2m-m_r$, $\delta_v=Z_2v/v_r-1$,
$\delta_1=Z_2Z_3^{1/2}e/e_r-1\equiv Z_1-1$,
$\delta_c=Z_1\beta/\beta_r-1$,
$\delta_{1\lambda}=Z_2^2\lambda_1-\lambda_{1r}$ and
$\delta_{2\lambda}=Z_2^2\lambda_2-\lambda_{2r}$ are determined by
renormalized conditions.

Since all the quantities in the following are renormalized ones, all
the subscripts $r$ will be dropped out.

\section { Calculation of VMM}
To compute the VMM we first show Feynman rules in Fig.
\ref{feynman} (a)-(e). The contribution to VMM up to $e^2$ is depicted
by Fig.\ref{feynman}(g), two diagrams in Fig.\ref{feynman}(h), which
is denoted by $\delta \Gamma_l^\mu$, $l=1,\,2$,
Fig.\ref{feynman}(i), which is denoted by $\Gamma^\mu$, and the
counterterm, Fig.\ref{feynman})f). The scattering amplitude of
carrier under external gauge field is \bee \ci\mathcal{M}=-\ci
e^\prime \{\bar{u}(p^\prime)(1+\delta_1)\gamma^\mu
u(p)+\Gamma^\mu+\sum\limits_{l=1}^2\delta
\Gamma_l^\mu\}A_{\mu}^{cl}(\vv{q}) , \ee where counterterm
$\delta_1$ plays a similar role as $Z_{int}-1$ and $Z_{kin}-1$ in
Ref. \cite{npb}, and $A_{\mu}^{cl}(\vv{q})$ is the Fourier
transformation of the external field, $p^\prime$ and $p$ are
outgoing and incoming momentums of carrier respectively. Since we
are working in lower energy limit, we ignore the renormalization of
Fermi velocity and charge. We get \beea \delta \Gamma_l^\mu&=&-\ci
\lambda_l\int \frac{d^3k}{(2\pi)^3} \frac{\bar{u}(p^\prime)\gamma^l
 (\sd{\tilde{k}}-m)\gamma^\mu
(\sd{\tilde{k}}^\prime-m)\gamma_l u(p)} {[\sd{\tilde{k}}^2-m^2+\ci
\eta][\sd{\tilde{k}}^{\prime 2}-m^2]} \nonumber \\
 \Gamma^\mu&=&\frac{2\pi\ci e^2}{\varepsilon v^4}
 \int\frac{d^3k}{(2\pi)^3} \frac{u^\dag(p^\prime)
 (\sd{\tilde{k}}^{\prime\prime}+m)\gamma^\mu
(\sd{\tilde{k}}+m)\gamma_0 u(p)} {[\sd{\tilde{k}}^{\prime\prime
2}-m^2+\ci\eta][\sd{\tilde{k}}^2-m^2]|\vv{p}-\vv{k}|} \nonumber
\\ && \label{gamma} \eea respectively, where $q=p^\prime-p$,
$k^{\prime\prime}=k+q$ and $k^\prime=k-q$. In the above equation, we
do not sum over the repeated index $l$ and the terms proportional to
$\beta^2\sim 10^{-4}$ are neglected.

\begin{center}
%~\\[\intextsep]
\begin{minipage}{0.45\textwidth}
\centering \vskip 0.2in
\includegraphics[width=2in]{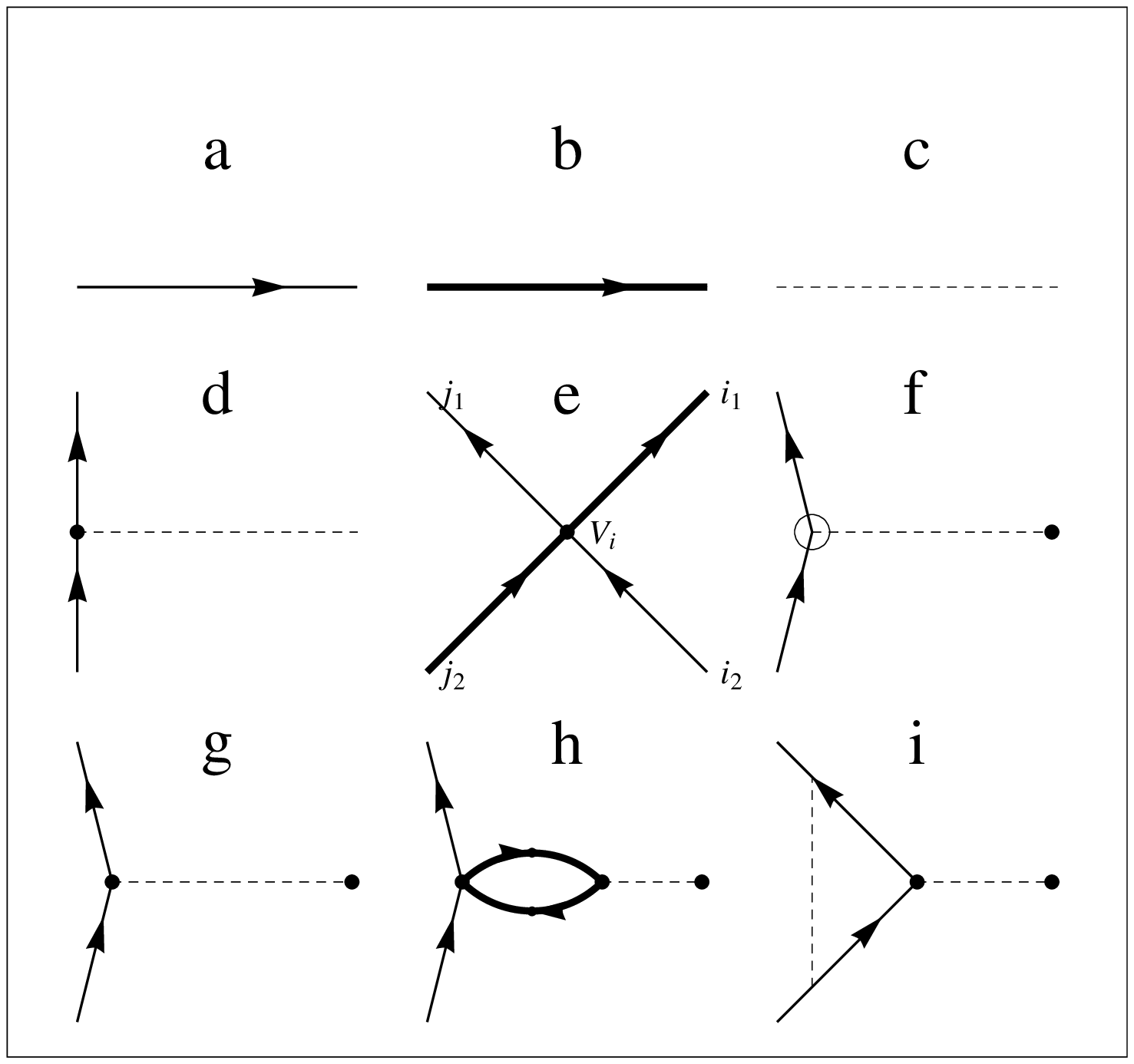}%
\vskip 1.9in \figcaption{Feynman rule and Feynman diagram on
graphene. a) Propagator of $\varphi$ field with momentum $p$,
$\ci\frac{\sdd{\tilde{p}}+m}{(\tilde{p})^2-m^2+\ci \epsilon}$, where
$\epsilon$ is infinitesimal positive. b) Propagator of $\chi$ field,
$\ci\frac{\sdd{\tilde{p}}-m}{(\tilde{p})^2-m^2+\ci \epsilon}$. c)
Propagator of gauge field, $\frac{2\pi\ci}{\varepsilon}\frac{
g_{\mu\nu}}{p-\ci \epsilon}$. d) Interaction vertex between
$\varphi$ field and gauge field, $-\ci e^\prime\gamma^{\mu}$, where
$e^\prime=e$ for $\mu=0$ and $e^\prime=\beta e$ for $\mu=1,\,2$. For
$\chi$ field the interaction is similar.  e) Two vertices of
four-fermion interactions, $-\ci
\lambda_1\gamma^1_{i_1i_2}\gamma_{1j_1j_2}$ and $-\ci
\lambda_2\gamma^2_{i_1i_2}\gamma_{2j_1j_2}$. f)Counterterm vertex,
$-\ci e^\prime \delta_1\gamma^{\mu}$. Since we are only concerned
about the correction to VMM up to order $e^2$, the renormalization
of fermion velocity is ignored. g) Tree level diagram contributing
to VMM. h) Four-fermion corrections. i) Vertex correction.
  } \label{feynman}
\end{minipage}
\setlength{\intextsep}{0.2in plus 0in minus 0.1in}
%\\[\intextsep]
\end{center}

Since the external field is time-independent, $q^0=p^{\prime
0}-p^0=0$ in Eq. (\ref{gamma}). If the electromagnetic field varies
very slowly over a large region, Fourier components of the
electromagnetic field will be concentrated about $\vv{q}=0$. We can
thus take nonrelativistic limit, $\vv{q}\rightarrow 0$, in $\ci
\mathcal{M}$, which means $|\vv{p}|,\,|\vv{p}^\prime|,\,|\vv{q}|\le
m/v$. Therefore, we have relations $-(\tilde{q})^2=v^2\vv{q}^2>0$
and $\tilde{p}\tilde{p}^\prime=
(p^0)^2-v^2\vv{p}\cdot\vv{p}^\prime\approx m^2$.

To study the response to external magnetic field, we set time
component of $A_{cl}$ as zero, {\it i.e.}
$A_{cl}(\vv{q})=(0,\vv{A}(\vv{q}))$. We therefore only need to
calculate the spatial part in $\ci\mathcal{M}$.

Since our theory violets the "Lorentz covariance", we should treat the
result carefully. Furthermore, all the integrals in Eq.
(\ref{gamma}) are divergent and therefore the result seems
ambiguous. However, we have the good news that the ambiguity have no
effect on VMM. After a lengthy calculation, such as Wick
rotation\ct{field-theory} and the expansion of the result to order
$|\vv{p}|,\,|\vv{p}^\prime|,\,|\vv{q}|$, we write $\Gamma^i$ and
$\delta\Gamma^i_l$ as $\bar{u}(C_1\gamma^i+C_2\ci
\epsilon^{ij}q^j\sigma_3/2)u$ in nonrelativistic limit, where
$\epsilon^{12}=-\epsilon^{21}=1$, $\epsilon^{11}=\epsilon^{22}=0$,
$C_1$ and $C_2$ depend on $\Gamma^\mu$, $\delta\Gamma^i_1$ and
$\delta\Gamma^i_2$. For all the cases, $C_1$ is divergent while
$C_2$ is finite. Together with the tree diagram and counterterm,
$(1+\delta_1+
C_1(\Gamma^i)+C_1(\delta\Gamma^i_1)+C_1(\delta\Gamma^i_2))\bar{u}\gamma^i
u$ should be fixed to match renormalization conditions. Comparing
with the Born approximation for scattering from a potential of
carrier nearly $\vv{p},\vv{q}\rightarrow 0$, we find that
$1+\delta_1+
C_1(\Gamma^i)+C_1(\delta\Gamma^i_1)+C_1(\delta\Gamma^i_2)$ is just
the electric charge of carrier, in units of $e$. Due to this, we set
the renormalization condition as $1+\delta_1+
C_1(\Gamma^i)+C_1(\delta\Gamma^i_1)+C_1(\delta\Gamma^i_2)=1$ at
$\vv{p},\vv{q}\rightarrow 0$. This renormalization condition
corresponds with the fact that the carrier at lower energy ($\vv{p}=0$) possesses unit (renormalized) charge $e$ when scattered
under external potential which varies very slowly.

For finite term $C_2$, we have \bee
C_2(\Gamma^i)=\frac{e^2}{4\varepsilon m},
\,~~C_2(\delta\Gamma^i_l)=-\frac{\lambda_l }{4\pi v}. \ee

Ignoring term proportional to $\vv{p}+\vv{p}^\prime$, which is the
contribution of the operator $\vv{p}\cdot\vv{A}+\vv{A}\cdot\vv{p}$
in the standard kinetic energy term of nonrelativistic quantum
mechanics, we rewrite $\bar{u}\gamma^iu$ term as \bee
\bar{u}(p^\prime)\gamma^i u(p)\rightarrow \frac{-\ci v \epsilon^{ij}
q^j}{m} \bar{u}\frac{\sigma^3}{2} u. \ee
 We obtain, then, \bee
\ci\mathcal{M}=-\ci 2m\xi^\dag\frac{\sigma_3}{2} \xi{ev\beta\over
m}(1-\frac{e^2}{4\varepsilon v}+\frac{m}{4\pi
v^2}(\lambda_1+\lambda_2))B^3,\ee where $B^3=-\ci
(q^1A_{cl}^2-q^2A_{cl}^1)$ is magnetic field perpendicular to
graphene, $\xi=(1,0)^T$ is two-component spinor and
$\xi^\dag\frac{\sigma_3}{ 2}\xi=1/2\equiv s_3$ indicates that
electron pseudo-spin is 1/2.

We interpret $\mathcal{M}$ as the Born approximation to the
scattering of the electron from a potential. The potential is just
that of a magnetic moment interaction, $V(\vv{x})=-\mu_e^3(\vv{K})
B^3$, where \bee \mu_e^3(\vv{K})=\frac{ev\beta}{ m}(1-\frac{e^2}{
4\varepsilon v}+\frac{m}{ 4\pi v^2}(\lambda_1+\lambda_2))
s_3\label{moment+k}\ee is the carrier VMM parallel to $B^3$ at
$\vv{K}$ valley. For the hole, we get the same value with a
necessary minus sign. Similarly, for carrier at $-\vv{K}$ valley,
VMM is also the same with a minus sign. Such phenomenon is known as
the broken inversion symmetry in Ref. \cite{niu}.

By recovering $\hbar$, the leading term of VMM is ${e\hbar
v\beta/m}$, which is also obtained by ref. \cite{niu}. However,
besides the leading term, there are also other contributions to VMM.
The relative contribution to VMM are \bee-\frac{e^2}{ 4\varepsilon
\hbar v}-\frac{mV_s}{ 4\pi\hbar^2 v^2}=\alpha(-\frac{1}{
4}+\frac{1.55\varepsilon\, m\,a}{4\pi\hbar v}),\label{relative}\ee
where $\alpha=e^2/(\varepsilon\hbar v)\approx 0.73$ when
$\varepsilon=3$. Substituting $a=2.46\overset{\circ}{A}$, $v\approx
10^{-8}cm/s$ into Eq. (\ref{relative}), we find the relative
modifications to VMM due to vertex correction and four-fermion
interactions are about $-18\%$ and $3\%$ respectively if
we choose $m=0.26eV$.

It looks strange that it is not $V_d$ but $V_s$ which contributes to
VMM. Such behavior stems from the definition of $\chi$ field. From
the definition of $\varphi$ field and $\chi$ field $V_d$ only
relates to interaction between carriers in different valley with the
same pseudo-spin so that it does not contribute to VMM. On the
contrary, $V_s$ relates to interaction between carriers in different
valley with the different pseudo-spin. Therefore, only $V_s$
contributes to VMM.

\section {Discussions}
In this paper we have discussed the carrier interactions. The study
reveals that besides the well known Coulomb repulsion between
carriers, there are four-fermion interactions between carriers in
different valleys. Since the interactions are short-range and
contacting ones, they depend on the atom collocation detail.
Therefore, the four-fermion interactions are type dependent. Our
study shows that one of the four-fermion interactions attracts
carriers in different valleys, which we believe to be helpful in
understanding the unusual superconduction effect in graphene.

We also compute VMM from the tree level diagram, the vertex
correction and the four-fermion interactions respectively. The
contribution from the tree level diagram agrees with the result
obtained in ref. \cite{niu}. The other two contributions counteract
each other and therefore the total contribution to VMM is about
$-15\%$ if we choose $m=0.26eV$ and $\varepsilon=3$.

The very high accurate measurement of spin magnetic moment is very
important, both from theoretical viewpoint and from practical one.
Similarly, our result on VMM is also significant to valleytronics in
graphene, especially to the future apparatus design based on
valleytronics. Our study also points out that, in close analogy to
the Zeeman split, the contribution to VMM induced by $V_s$ is
inherent, since it is independent on the energy gap $m$. In other
words, to measure the magnetic moment induced by $V_s$, we may
choose the substrate freely, although different substrate may
generate different energy gap and different effective permittivity.

From Eq. (\ref{relative}), $\alpha$ plays the same role as the fine
structure constant in QED, $\alpha_e=\frac{e^2}{\hbar c}\approx
1/137$. However, because $\alpha$ is about 100 times larger than
$\alpha_e$, it is hard to state that we mimic QED by carrier
interaction in graphene. Meanwhile, when we deal with problems
dominated by quantum chromodynamics(QCD), especially in heavy
quarkonium, such as $c\bar{c}$ system and $b\bar{b}$ system, we
always take $\alpha_s=\frac{g^2_s}{\hbar c}$, where $g_s$ is the QCD
coupling, as the estimate of the effectiveness of perturbational
expansion. (In many cases when we deal with such problem we take an
approach very similar to QED, up to an unimportant color factor.) At
energy scale $740Mev$, $\alpha_s(740Mev) \approx 0.73\approx
\alpha$\ct{field-theory}. Noticing that the energy scale is close to
the soft scale of $c\bar{c}$ and $b\bar{b}$ systems\ct{0702105}, the
dynamics of which is depicted by nonrelativistic QCD, we conclude
that we can mimic the heavy quarkonium system by carrier
interactions in graphene. Therefore, the study on the heavy-quarkonium
system can also be carried out in graphene.

This work is supported by the Cultivation Fund of the
Key Scientific and Technical Innovation Project-Ministry
of Education of China (No. 708082).

\balance

\end{multicols}
\end{document}